\documentclass[dvips]{article}
\usepackage{icrctc07}

\title{Search for a Dark Matter annihilation signal from the Sagittarius dwarf galaxy with H.E.S.S.}
\shorttitle{Search for a Dark Matter annihilation signal from the
Sagittarius dwarf galaxy}
\authors{E. Moulin$^{1}$, C. Farnier$^{2}$, J.-F. Glicenstein$^{1}$, A. Jacholkowska$^{2}$, L. Rolland$^{3}$, M. Vivier$^{1}$, for the H.E.S.S. Collaboration}
\shortauthors{E. Moulin and et al}
\afiliations{$^1$DAPNIA/DSM/CEA, CE Saclay, F-91191 Gif-sur-Yvette Cedex, France\\
$^2$ Laboratoire de Physique Th\'eorique et Astroparticules,
Universit\'e Montpellier II, CC 70, Place Eug\`ene Bataillon, F-34095 Montpellier Cedex 5, France\\
$^3$ Laboratoire d'Annecy-le-Vieux de Physique des Particules,
IN2P3/CNRS, 9 Chemin de Bellevue, BP 110 F-74941 Annecy-le-Vieux
Cedex, France} \email{emmanuel.moulin@cea.fr}

\abstract{Dwarf Spheroidal galaxies are amongst the best targets
to search for a Dark Matter (DM) annihilation signal. The
annihilation of WIMPs in the center of Sagittarius dwarf
spheroidal (Sgr dSph) galaxy would produce high energy
$\gamma$-rays in the final state. Observations carried out with
the H.E.S.S. array of Imaging Atmospheric Cherenkov telescopes are
presented. A careful modelling of the Dark Matter halo profile of
Sgr dwarf was performed using latest measurements on its
structural parameters. Constraints on the velocity-weighted cross
section of Dark Matter particles are derived in the framework of
Supersymmetric and Kaluza-Klein models.
}

\begin{document}
\maketitle
%Begin the section.

\section{Introduction}
Astrophysical and cosmological observations provide a substantial
body of evidences for the existence of Cold Dark Matter (CDM)
although its nature remains still unknown. It is commonly assumed
that CDM is composed of yet undiscovered non-baryonic particles
for which plausible candidates are Weakly Interacting Massive
Particles (WIMPs). In most theories, candidates for CDM are
predicted in theories beyond the Standard Model of particle
physics~\cite{bertone}. The annihilation of WIMPs into
$\gamma$-rays may lead to detectable very high energy (VHE, E $>$
100 GeV) $\gamma$-ray fluxes above background via continuum
emission from the hadronization and decay of the cascading
annihilation products, predominantly from $\pi^0$'s generated in
the quark jets. Among the best-motivated CDM candidates are the
lightest neutralino $\tilde{\chi}$ provided by R-parity conserving
supersymmetric extensions of the Standard Model~\cite{jungman},
and
the lightest Kaluza-Klein particle (LKP)~\cite{servant} %provided
%by the
in universal extra dimension theories which is most often the
first KK
mode of the hypercharge gauge boson, $\tilde{B}^{(1)}$.
% is the
%best motivated.
% hadronization
%of gauge bosons and heavy quarks, or $\gamma$-ray lines through
%loop-induced processes.
%The annihilation of DM particles can generate $\gamma$-ray fluxes
%through different processes depending on the particle physics
%scenarios.Generally, WIMP annihilations will produce a continuum
%of $\gamma$-rays with energies up to the WIMP mass issued from the
%hadronization and decay of the cascading annihilation products,
%predominantly from $\pi^0$'s generated in the quark jets. Among
%the best motivated CDM candidates are the lightest neutralino
%$\tilde{\chi}$ provided by R-parity conserving supersymmetric
%extensions of the Standard Model~\cite{jungman}, and the lightest
%Kaluza-Klein particle (LKP)~\cite{servant} provided by the
%universal extra dimension (UED) theories in which the first KK
%mode of the hypercharge gauge boson, $\tilde{B}^{(1)}$ is the best
%motivated.\\
The H.E.S.S. array of Imaging Atmospheric Cherenkov Telescopes
(IACTs), designed for high sensitivity measurements in the 100 GeV
- 10 TeV energy regime, is a suitable instrument to
detect VHE $\gamma$-rays and investigate their possible origin.\\
Dwarf Spheroidal galaxies such as Sagittarius or Canis Major,
discovered recently in the Local Group, are among the most extreme
DM-dominated environments. Indeed, measurements of roughly
constant radial velocity dispersion of stars imply large mass to
luminosity ratios~\cite{wilkinson}.
%Nearby dwarfs are ideal
%astrophysical probes of the nature of DM as they usually consist
%of a stellar population with no hot or warm gas, no cosmic ray
%population and little dust.
The core of the Sgr dSph at l=5.6$^{\circ}$ and b=-14$^{\circ}$ in
galactic coordinates at a distance of about 24 kpc from the
Sun~\cite{majewski}. Latest velocity dispersion measurements on M
giant stars with 2MASS yields a mass to light ratio of about
25~\cite{majewski2}.
%The Sgr
%dSph core is positioned behind the bulge of Milky Way but outside
%the Galactic plane, thus reduced foreground  $\gamma$-ray
%contaminations are expected.
The luminous density profile of Sgr
dSph has two components~\cite{monaco}. The compact component,
namely the core, is characterized by a size of about 3 pc FWHM,
which corresponds to a point-like region for H.E.S.S. This is the
DM annihilation region from which $\gamma$-ray signal may be
expected. A diffuse component is well fitted by a King model with
a characteristic size of 1.6 kpc.\\
We present in this paper the observations of the Sgr dSph galaxy
by the H.E.S.S. array of Imaging Atmospheric Cherenkov Telescopes.
A careful modeling of the Dark Matter halo using the latest
measurements on the structural parameters of Sagittarius is
presented to derive constraints on the WIMP velocity-weighted
annihilation rate.

\section{Search for VHE $\gamma$-rays from observations of Sagittarius dwarf by H.E.S.S.}
H.E.S.S. (High Energy Stereoscopic System)
%is an array of four
%imaging atmospheric Cherenkov telescopes located in the Khomas
%Highlands of Namibia at an altitude of 1800 m a.s.l. Each
%telescope consists of an optical reflector of about 107 m$^2$
%effective area which collects the Cherenkov light emitted by the
%charged particles composing the electromagnetic shower initiated
%by the interaction of the primary $\gamma$-ray in the Earth's
%upper atmosphere. The light is focused on a 960 fast
%photomultiplier (PMT) camera~\cite{vincent}. %The total field of
%%view of the H.E.S.S. instrument is 5$^{\circ}$ in diameter.
%The stereoscopy technique used in the imaging atmospheric
%Cherenkov telescopes allows for accurate reconstruction of the
%direction and energy of the primary $\gamma$-rays as well as an
%efficient rejection of the background induced by cosmic ray
%interactions~\cite{trigger}. The energy threshold of H.E.S.S. at
%zenith before selection cuts is 160 GeV due to the degradation of
%the optical performance in 2006. The point source sensitivity is
%better than $\rm 2\times 10^{-13} cm^{-2}s^{-1}$ above 1 TeV for a
%5$\sigma$ detection in 25 hours~\cite{crabe}.\\
has observed the Sgr dSph in June 2006 with zenith angles ranging
from 7$^{\circ}$ to 43$^{\circ}$ around an average value of
19$^{\circ}$. A total of 11 hours of high quality data are
available for the analysis after standard selection cuts. After
calibration of the raw shower images from PMT
signals~\cite{aharonian0}, two independent reconstruction
techniques were combined to select $\gamma$-ray events and
reconstruct their direction and energy. The first one uses the
Hillas moment method~\cite{aharonian1}. The second analysis
referred hereafter as ``Model Analysis'', is based on the
pixel-per-pixel comparison of the shower image with a template
generated by a semi-analytical shower development
%The event
%reconstruction relies on a maximum likelihood method which uses
%available pixels in the camera, without requirement for an image
%cleaning
model~\cite{denauroi0,denauroi1}.
%The reconstructed shower
%parameter (energy, impact, direction and primary interaction
%point) are obtained as a product of the minimization procedure.
 The separation between $\gamma$ candidates and hadrons is done
 using a combination of the Model goodness-of-fit
 parameter~\cite{denauroi1} and the Hillas mean scaled width and length
 parameters, which results in an improved background
 rejection~\cite{crabe}.
%Standard cuts on the width and the length of Hillas ellipses
%combined with the goodness-of-fit are used to suppress the
%hadronic background~\cite{aharonian1}.
An additional cut on the primary interaction depth is used to
improve background rejection.\\
%Both methods yields typical energy resolution of 15\% above energy
%threshold.
%The analysis yields a typical energy resolution of 15\%
%above energy threshold and the angular resolution at the 68\%
%containment radius is found to be better than 0.06$^{\circ}$
%per $\gamma$-ray.\\
The on-source signal is defined by integrating all the events with
angular position $\theta$ in a circle around the target position
with a radius of $\theta_{cut}$. The target position is chosen
according to the photometric measurements of the Sgr dSph luminous
cusp showing that the position of the center corresponds to the
center of the globular cluster M 54~\cite{monaco1}. The target
position is thus found to be $\rm (RA = 18^h55^m59.9s, Dec =
-30d28'59.9'')$ in equatorial coordinates (J2000.0).
%or $\rm (l =
%5^{\circ}41'12.9'',b = -14^{\circ}16'29.8'')$ in Galactic
%coordinates.
The signal coming from Sgr dSph is expected to come from a region
of 1.5 pc, about 30'', much smaller the H.E.S.S. point spread
function (PSF). A $\theta_{cut}$ value of 0.14$^{\circ}$ suitable
for a point-like source was therefore used in the analysis. In
case of a Navarro-Frenk-White (NFW) density profile~\cite{nfw} for
which $\rho$ follows r$^{-1}$ or a cored profile~\cite{evans}
folded with the point spread function (PSF) of H.E.S.S., the
integration region allows to retrieve a significant fraction of
the expected signal. See Table~\ref{tab:table}.\\
%A cut on the image size of 60
%photoelectrons is used to obtain a good sensitivity for weak
%sources. In order to reduce systematic effects which affect images
%close to the edges of the camera, only events reconstructed within
%a maximum distance of 2.5$^{\circ}$ from the camera center are
%used for this analysis. The excess sky map is obtained by the
%subtraction of a background model on the $\gamma$-ray candidate
%sky distribution. The background level is estimated using the
%ring-background method~\cite{puhlhofer} where the background rate
%is calculated from the integration of $\gamma$-like events falling
%in an annulus around the center of the camera with identical
%observation conditions and acceptances than that used for the
%on-source region, which allows
%an estimate of the background on every sky position.\\
No significant $\gamma$-ray excess is detected in the sky map. We
thus derived the 95\% confidence level upper limit on the observed
number of $\gamma$-rays: $\rm N_{\gamma}^{95\%\, C.L.}$. The limit
is computed knowing the numbers of events in the signal and
background regions above the energy of 250 GeV
%in the signal region
%$\rm N_{ON} = 437$, in the background region $\rm N_{OFF} = 4270$,
%and the ratio of the off-source time to the on-source time
%$\rm\alpha = 10.1$. We
using the Feldman \& Cousins method~\cite{feldman} and we obtain:
%\begin{equation}
$\rm N_{\gamma}^{95\%\,C.L.} = 56$.
%\end{equation}
Given the acceptance of the detector for the observations of the
Sgr dSph, a 95\% confidence level upper limit on the $\gamma$-ray
flux is also derived:
\begin{eqnarray}
\Phi_{\gamma}({\rm E_{\gamma} > 250\,GeV}) < 3.6 \times 10^{-12}
\,{\rm cm^{-2}s^{-1}}\, \nonumber \\(95\%\,C.L.)
\end{eqnarray}

\section{Predictions of $\gamma$-ray from Dark Matter annihilations}
\begin{figure}[!bp]
\begin{center}
\includegraphics [width=0.48\textwidth]{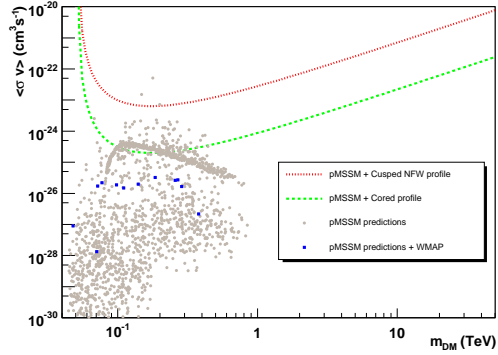}
\end{center}
\caption{Upper limits at 95\% C.L. on $\langle \sigma v \rangle$
versus the DM particle mass in the case of a cusped NFW (red
dotted line) and a cored (green dashed line) DM halo profiles
respectively. The pMSSM parameter space was explored with DarkSUSY
4.1~\cite{darksusy}, each point on the plot corresponding to a
specific model (grey point). Amongst these models, those
satisfying in addition the WMAP constraints on the CDM relic
density are overlaid as blue square. The limits in case of
neutralino dark matter from pMSSM are derived using the
parametrisation from~\cite{bergstrom} for a higgsino-type
neutralino annihilation $\gamma$ profiles.}\label{fig:sigmav}
\end{figure}

The $\gamma$-ray flux from  annihilations of DM particles of mass
$m_{DM}$ accumulating in a spherical DM halo can be expressed in
the form:
\begin{equation}\label{equ4}
\frac{d\Phi(\Delta\Omega,E_{\gamma})}{dE_{\gamma}}\,=\frac{1}{4\pi}\,\underbrace{\frac{\langle
\sigma
v\rangle}{m^2_{DM}}\,\frac{dN_{\gamma}}{dE_{\gamma}}}_{Particle\,
Physics}\,\times\,\underbrace{\bar{J}(\Delta\Omega)\Delta\Omega}_{Astrophysics}
\end{equation}
as a product of %characterized by
a particle physics component with an astrophysics component. The
particle physics part contains $\langle \sigma v\rangle$, the
velocity-weighted annihilation cross section, and
$dN_{\gamma}/dE_{\gamma}$,  the differential $\gamma$-ray spectrum
summed over the whole final states with their corresponding
branching ratios.
\begin{table*}[!htp]
\begin{center}
\begin{tabular}{|c||c|c|c|}
\hline
Halo type&Parameters&$\rm \bar{J}$&Fraction of signal \\
&&$\rm (10^{24} GeV^{2}cm^{-5})$&in $\rm\Delta\Omega = 2 \times 10^{-5}$sr\\
%&$r_c/r_s$&$v_a$/A&$\bar{J}$\\
\hline \hline
%Cusped halo ($\rm \gamma = 1$)&r$_s$ = 0.2 kpc&$\rm 2.2$& \\
Cusped NFW halo &r$_s$ = 0.2 kpc&$\rm 2.2$& 93.6\%\\
&$\rm A = \rm 3.3\times 10^7 M_{\odot}$&& \\
\hline
Cored halo&r$_c$ = 1.5 pc&$\rm 75.0$&99.9\%\\
&v$_a$ = 13.4 $\rm km\ s^{-1}$&&\\
\hline
\end{tabular}
\end{center}
\caption{\label{tab:table}Structural parameters for a cusped NFW
($\rm r_s,A$) and a cored ($\rm r_c,v_a$) DM halo model,
respectively. The values of the solid-angle-averaged l.o.s
integrated squared DM distribution are reported in both cases for
the solid angle integration region $\rm \Delta \Omega = 2\times
10^{-5} sr$.}
\end{table*}
The astrophysical part corresponds to the line-of-sight-integrated
squared density of the DM distribution J, averaged over the
instrument solid angle integration region for H.E.S.S. ($\rm
\Delta\Omega=2\times 10^{-5}$sr):
\begin{eqnarray}
\label{eqnj}
%J\,=\,\int_{l.o.s}\rho^2(r[s])ds \nonumber \\
\bar{J}(\Delta\Omega)\,=\,\frac{1}{\Delta\Omega}\int_{\Delta\Omega}
{\rm PSF}*\int_{l.o.s}\rho^2(r[s])ds\,d\Omega
\end{eqnarray}
where PSF is the point spread function of H.E.S.S.\\
The mass distribution of the DM halo of Sgr dwarf has been
described by plausible models taking into account the best
available measurements of the Sgr dwarf galactic structure
parameters. We have used two widely different models. The first
has a NFW cusped profile~\cite{nfw} with the mass density given
by:
\begin{equation}
\rho_{NFW}(r)\,=\,\frac{A}{r(r+r_s)^{2}}
\end{equation}
with A the normalization factor and $r_{s}$ the scale radius taken
from~\cite{evans}. Using Eq.~\ref{eqnj}, the value of $\bar{J}$
obtained with this model is reported in Table~\ref{tab:table}.
%In the spirit of reference~\cite{evans},
We have also studied a core-type halo model  as in~\cite{evans}
characterized by the mass density:
\begin{equation}
\rho_{core}(r)\,=\,\frac{v_a^2}{4\pi G}\frac{3 r_c^2
+r^2}{(r_c^2+r^2)^2}
\end{equation}
where $r_c$ is the core radius and $v_a$ a velocity scale.
 However, we have tried to update the $v_{a}$ and $r_{c}$
values which were used in~\cite{evans}. By inserting in the Jeans
equation the luminosity profile of the Sgr dwarf core of the form:
\begin{equation}
\nu (r)= \frac{\nu_{0}{r_c}^{2\alpha}}{(r_c^2+r^2)^{\alpha}}
\end{equation}
we estimated from the data of \cite{monaco1} $\alpha = 2.69\pm
0.10$ and $ r_c = 1.5\ \mbox{pc}.$ Note that the value of $r_{c}$
is only an upper limit. The value of the central velocity
dispersion of Sgr Dwarf is
 $\sigma = 8.2\,\pm\, 0.3\,\rm km\,s^{-1}$~\cite{zaggia}.
We have assumed that the velocity dispersion is independent of
position. The value of $v_{a}$ is then given by $v_{a} =
\sqrt{\alpha}\,\sigma = 13.4\,\rm km\, s^{-1}$. The cored model
gives a very large value of $\bar{J},$ which is reported in
Table~\ref{tab:table}. The third column of Table~\ref{tab:table}
gives the amount of signal expected in the solid angle integration
region $\rm \Delta\Omega = 2\times 10^{-5}$sr.
Fig.~\ref{fig:sigmav} shows the limits in the case of a cored
(green dashed line) and cusped NFW (red dotted line) profile using
the value of $\bar{J}$ computed above.
%The differential
%$\gamma$ spectrum is parametrized using the expression given
%in~\cite{bergstrom} for a higgsino-type neutralino.
Predictions for phenomenological MSSM (pMSSM) models are displayed
(grey points) as well as those satisfying in addition the WMAP
constraints on the CDM relic density (blue points). The SUSY
models are calculated with DarkSUSY4.1~\cite{darksusy}.
% in pMSSM
%framework and characterized by a basic set of independent
%parameters  : the higgsino mass parameter $\mu$, the gaugino mass
%parameter M$_2$, the CP-odd Higgs mass M$_A$, the common scalar
%mass m$_0$ , the trilinear couplings A$_{t,b}$ and the Higgs
%vacuum expectation value ratio $\tan \beta$. The set of parameters
%for a given model is randomly chosen in a parameter region
%encompassing a large class of pMSSM models, as described in
%Tab.~\ref{tab:table2}.
In the case of a cusped NFW profile, the
H.E.S.S. observations do not establish severe constraints on the
velocity-weighted cross section. For a cored profile, due to a
higher central density,
stronger constraints are derived and some pMSSM models
%yielding neutralino masses in the 100 - 400 GeV range,
can be excluded in the upper part of the pMSSM scanned region. \\
In the case of KK dark matter, the differential $\gamma$ spectrum
is parametrized using Pythia~\cite{pythia} simulations and
branching ratios from~\cite{servant}. Predictions for the
velocity-weighted cross section of B$^{(1)}$ dark matter particle
are performed using the formula given in~\cite{sigmaKK}. In this
case, the expression for $\langle \sigma v \rangle$ depends
\begin{figure}[!hb]
\begin{center}
\includegraphics [width=0.48\textwidth]{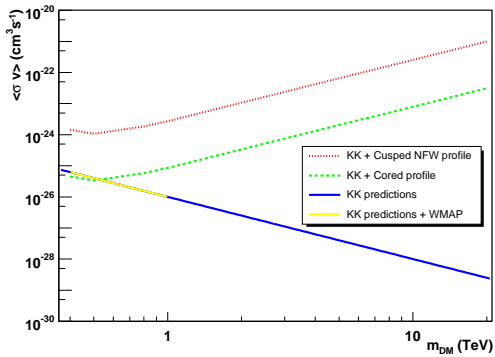}
\end{center}
\caption{Upper limits at 95\% C.L. on $\langle \sigma v \rangle$
versus the DM particle mass in the  B$^{(1)}$ Kaluza-Klein
scenarios for a cusped NFW (red dotted line) and a cored (green
dashed line) DM halo profiles respectively. The blue line
corresponds to Kaluza-Klein models~\cite{servant}. Overlaid
(yellow line) are the KK models satisfying WMAP constraints on the
CDM relic density.}\label{fig:sigmav2}
\end{figure}
analytically on the $\rm \tilde{B^{(1)}}$ mass square.
Fig.~\ref{fig:sigmav2} shows the sensitivity of H.E.S.S. in the
case of Kaluza-Klein models where the hypercharge boson B$^{(1)}$
is the LKP, for a cored (green solid line) and a cusped NFW (red
solid line) profile respectively using the value of $\rm \bar{J}$
computed in section 3.2. With a NFW profile, no Kaluza-Klein
models can be tested. In the case of a cored model, some models
providing a LKP relic density compatible with WMAP contraints can
be excluded. From the sensitivity of H.E.S.S., we exclude
B$^{(1)}$ masses lying in the range 300 - 500 GeV.
%derive a lower limit on the B$^{(1)}$
%mass of 500 GeV.

\section{Conclusions}
The observations of Sgr dSph with H.E.S.S. reveal no significant
$\gamma$-ray excess at the nominal target position. The
Sagittarius dwarf DM halo profile has been modeled using latest
measurements of its structure parameters. Constraints have been
derived on the velocity-weighted cross section of the DM particle
in the framework of supersymmetric and Kaluza-Klein models.

%\section{Acknowledgements}
%The support of the Namibian authorities and of the University of
%Namibia in facilitating the construction and operation of H.E.S.S.
%is gratefully acknowledged, as is the support by the German
%Ministry for Education and Research (BMBF), the Max Planck
%Society, the French Ministry for Research, the CNRS-IN2P3 and the
%Astroparticle Interdisciplinary Programme of the CNRS, the U.K.
%Particle Physics and Astronomy Research Council (PPARC), the IPNP
%of the Charles University, the South African Department of Science
%and Technology and National Research Foundation, and by the
%University of Namibia. We appreciate the excellent work of the
%technical support staff in Berlin, Durham, Hamburg, Heildelberg,
%Palaiseau, Paris, Saclay, and in Namibia in the construction and
%operation of the equipment.
%\nocite{ref4} \nocite{ref5}
%\nocite{ref6} \nocite{ref7}
%This is the reference to .bib file (Whitout .bib!)
\bibliography{icrc0496}
%This in the bibtex style, is ok.
\bibliographystyle{plain}

\end{document}